\documentclass[%
 reprint,
 amsmath,amssymb,
 aps,
pra,
]{revtex4-1}

\usepackage{physics}
\usepackage{subcaption}
\usepackage{hyperref}
\usepackage{cleveref}
\usepackage{color}
\usepackage{graphicx}
\usepackage{dcolumn}
\usepackage{bm}
\usepackage{epstopdf}

\begin{document}

\preprint{APS/123-QED}

\title{Vibrational ladder-descending photostabilization of a weakly bound molecule: Quantum optimal control with a genetic algorithm}

\author{Mateo Londoño}
  \email{mateo.londono@correounivalle.edu.co}
\affiliation{%
 Departamento de Física, Universidad del Valle, A.A. 25360, Cali, Colombia}
 \author{Julio C. Arce}%
 \email{julio.arce@correounivalle.edu.co}
\affiliation{%
 Departamento de Química, Universidad del Valle, A.A. 25360, Cali, Colombia}%

\date{\today}

\begin{abstract}

We propose an optical control scheme for driving a polar diatomic molecule from a high-lying vibrational level to a target low-lying one, within the same electronic state.
The scheme utilizes an infrared chirped laser pulse with an analytical shape, whose parameters are optimized by means of a heuristic formulation of quantum optimal control based on a genetic algorithm.
We illustrate this methodology computationally for a KRb Feshbach molecule in the lowest triplet electronic state.

\end{abstract}

\maketitle


\section{\label{sec:intro}Introduction}

The formation of cold (1 mK$<T<$1 K) and ultracold ($T<$1 mK) ensembles of diatomic molecules in a controlled fashion \cite{Quemener2012,Koch2018} is a current challenge of great interdisciplinary interest \cite{Carr2009,Chin2009,Cote2014,Perez2020}.
Proposals have been put forth for the creation of such ensembles from the binary collisions in cold or ultracold atomic gases by one-step photoassociation (PA) \cite{Juarros2006,Kotochigova2007,Marquetand2007,Kallush2008,Molano2019}, two-step PA \cite{Ulmanis2012}, a combination of one-step and two-step PA \cite{deLima2017}, magnetoassociation \cite{Kohler2006}, and electroassociation \cite{Castano2020}.
One-step PA, magnetoassociation, and electroassociation involve only one Born-Oppenheimer potential energy curve (PEC), whereas two-step PA involves several PECs.

In all these association methods the molecules are typically left in a distribution of rovibrational levels of the ground and/or an excited electronic state. Hence, to achieve cooling, control schemes must be applied for the subsequent stabilization into low-lying rovibrational levels of the ground electronic state, including the absolute ground state.
For the situation where the molecules are left in high- or intermediate-lying levels of the ground electronic state, and if they are polar, controlled stabilization schemes within the same electronic state have been devised \cite{Marquetand2007,Ndong2010,deLima2015,Niu2018,Devolder2021}.
Of particular interest for this paper are those that entail a consecutive descent across the ladder of vibrational levels, using a single chirped laser pulse. For example, Marquetand and Engel employed local control theory to achieve one-step PA together with some stabilization during H+F and H+I collisions \cite{Marquetand2007}, and Devolder \emph{et al.} applied a quantum optimal control (QOC) method for the stabilization of a RbSr molecule formed previously via one-step PA \cite{Devolder2021}. However, the resulting optimal pulse has a complicated structure, making it very difficult to achieve experimentally.
Regardless of whether the molecules are left in the ground or an excited electronic state, it has been demonstrated that pump-dump \cite{Sage2005,Guerrero2018} and STIRAP \cite{Aikawa2010,Borsalino2014,Devolder2021} methodologies can achieve stabilization.
Nevertheless, both methodologies involve intermediate excited electronic states, which can introduce complications, like  internal conversions, intersystem crossings, and fast radiative decay into other electronic states.
In addition, it may be the case that the populated bound levels of the initial electronic state have relatively small Franck-Condon factors for the transitions to the vibrational levels of the intermediate excited electronic states \cite{Guerrero2018}. In this situation, it would be convenient to introduce a prior step to drive the molecules to the levels of the initial electronic state with the highest Franck-Condon factors.

In this paper, we address the problem of driving a polar diatomic molecule from an initial level into a target level of the same electronic state. Specifically, we demonstrate that QOC based on genetic algorithms (GAs) is an attractive alternative to accomplish vibrational ladder descending (LD), employing a linear chirped pulse (LCP) with an analytical shape that can be achieved experimentally with relative ease.
This scheme can be utilized
in two cases: either as a \emph{final} step for achieving further stabilization, after application of any of the association methodologies mentioned in the first paragraph of this Introduction, or as a \emph{prior} step in pump-dump or STIRAP methodologies for maximizing the Franck-Condon factors.
In Sec. \ref{sec:LD} we explain our LD scheme in the context of the second case. Specifically, we apply it to the model of Ref. \citenum{Guerrero2018}, where driving a weakly bound KRb Feshbach molecule in the lowest triplet electronic state to a lower-lying vibrational level is a desirable prior step to optimize a subsequent pump-dump stabilization method.
In Sec. \ref{sec:method} we briefly describe the numerical methods we used for solving the time-independent and time-dependent Schr\"odinger equations, and the QOC+GA methodology we employed for the optimization of the analytical pulse shape.
In Secs. \ref{subsec:OLD} and \ref{subsec:MLD} we present and discuss the simulation results for one-rung-at-a-time (OLD) and multiple-rung-at-a-time (MLD) variants of LD, respectively.
In addition, at the end of Sec. \ref{subsec:MLD} we comment on the current experimental feasibility of our proposal, given the state of the art in the generation of ultrashort laser pulses in the mid-to-far infrared domain.
Finally, in Sec. \ref{sec:conclusions} we state the conclusions of this work and suggest some perspectives for future developments.

\section{\label{sec:LD}The ladder-descending scheme}

\begin{figure}[]
\includegraphics[scale=0.36]{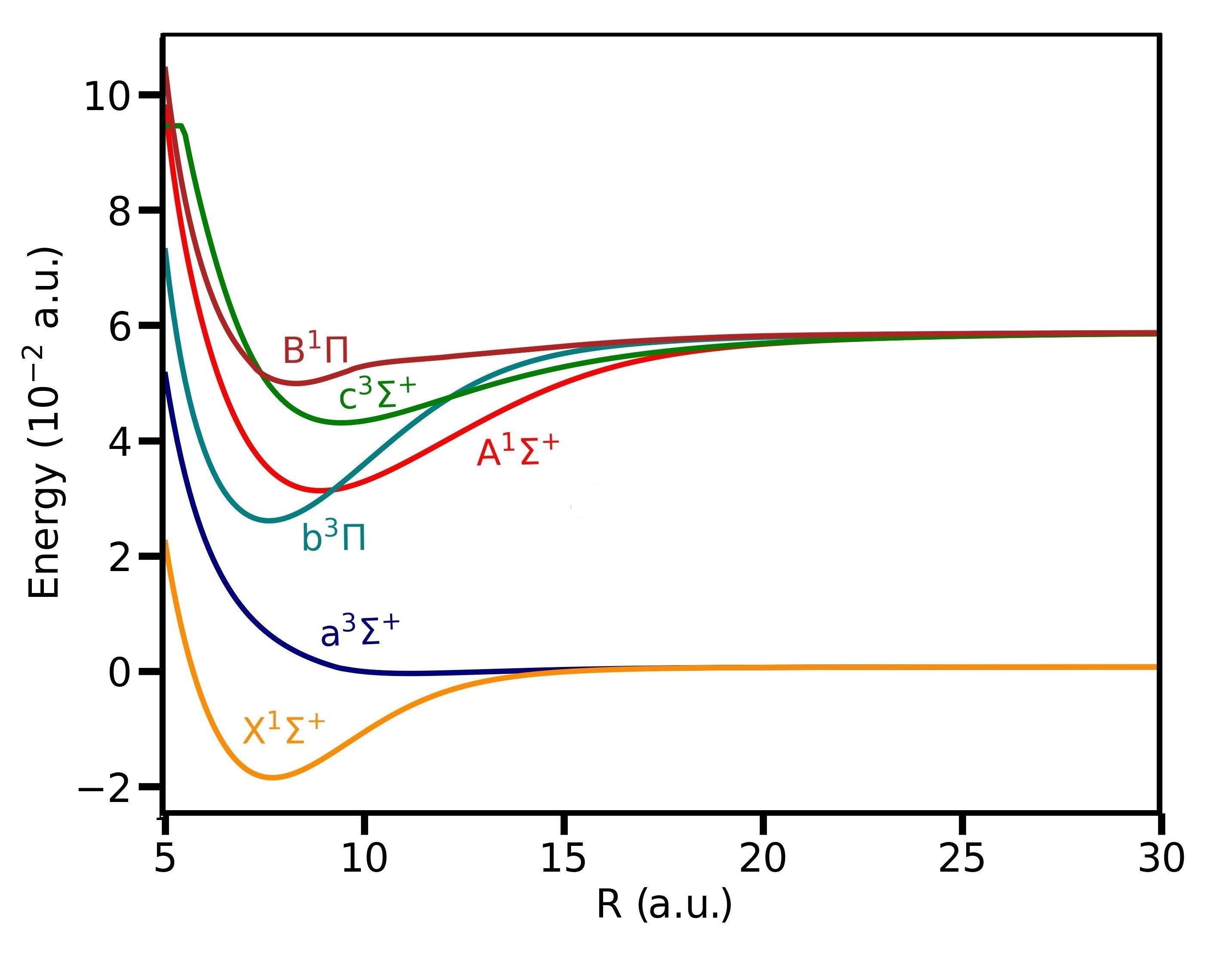}
\caption{(Color online) Potential energy curves for the KRb molecule.}
\label{fig:PECS_scheme}
\end{figure}

Arango and coworkers \cite{Guerrero2018} implemented a pump-dump scheme for the vibrational stabilization into the electronic ground state, $X^{1}\Sigma^{+}$, of a model $^{39}$K$^{87}$Rb Feshbach molecule formed in the lowest electronic triplet state, $a^{3}\Sigma^{+}$, using the $[b-A]$ scheme that involves the spin-orbit-coupled intermediate electronic states $b^{3}\Pi$ and $A^{1}\Sigma^{+}$ \cite{Borsalino2014}. The PECs corresponding to these electronic states are illustrated in Fig. \ref{fig:PECS_scheme}.
The pump pulse stimulates the $b^{3}\Pi$ $\leftarrow$ $a^{3}\Sigma^{+}$ absorption and the dump pulse stimulates the $A^{1}\Sigma^{+}$ $\rightarrow$ $X^{1}\Sigma^{+}$ emission.
Gaussian LCPs optimized by means of a GA were employed, either without (direct mechanism) or with (assisted mechanism) explicit consideration of the dynamics of the $b^{3}\Pi$ $\rightarrow$ $A^{1}\Sigma^{+}$ spin-orbit-induced radiationless transition.
The molecule was assumed to be initially in high-lying vibrational levels of the $a^{3}\Sigma^{+}$ state, where it may be formed through Feshbach tuning.
Unfortunately, for driving the molecule into its absolute ground state, $\ket{X^{1}\Sigma^{+},\upsilon=0}$, using this scheme such levels are far from optimal, since their (inter-curve) dipole couplings with the vibrational levels of the intermediate $b^{3}\Pi$ state are weak.
It turns out that the strongest inter-curve couplings occur for the $\upsilon=10$ level.
Hence, it is desirable to drive the molecule down to this level before applying the pump pulse.
However, a direct transition is not feasible, due to the (intra-curve) dipole coupling between well-separated levels being too weak.
In this paper we illustrate our infrared LD scheme by addressing this issue. Specifically, we show that this scheme allows bringing the molecule from any of the initial levels $\ket{a^{3}\Sigma^{+},\upsilon=20,24}$ to the optimal level  $\ket{a^{3}\Sigma^{+},\upsilon=10}$.

\begin{figure}[]
\includegraphics[scale=0.30]{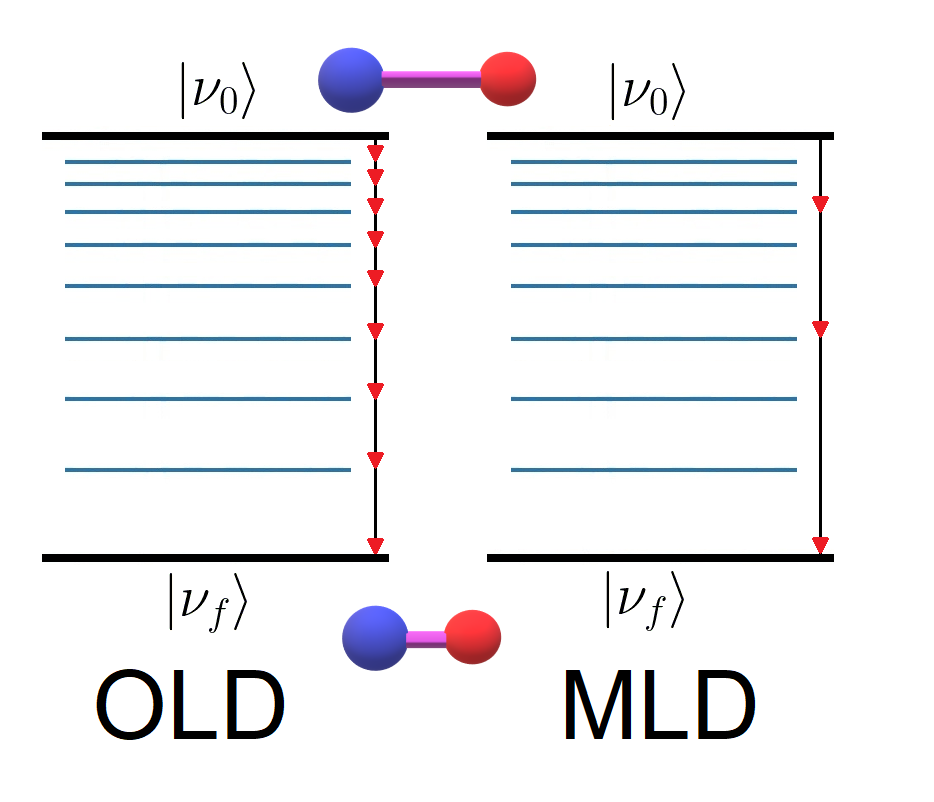}
\caption{(Color online) Schematic illustration of the one-rung-at-a-time (left) and multiple-rung-at-a-time (right) ladder-descending schemes.}
\label{fig:LD_schemes}
\end{figure}

\begin{figure}[]
\begin{subfigure}[h]{1.0\linewidth}
\includegraphics[scale=0.49]{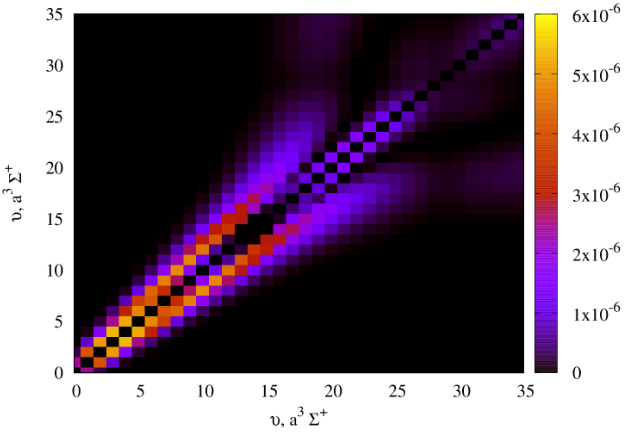}
\caption{}
\end{subfigure}
\begin{subfigure}[h]{1.0\linewidth}
\includegraphics[scale=0.28]{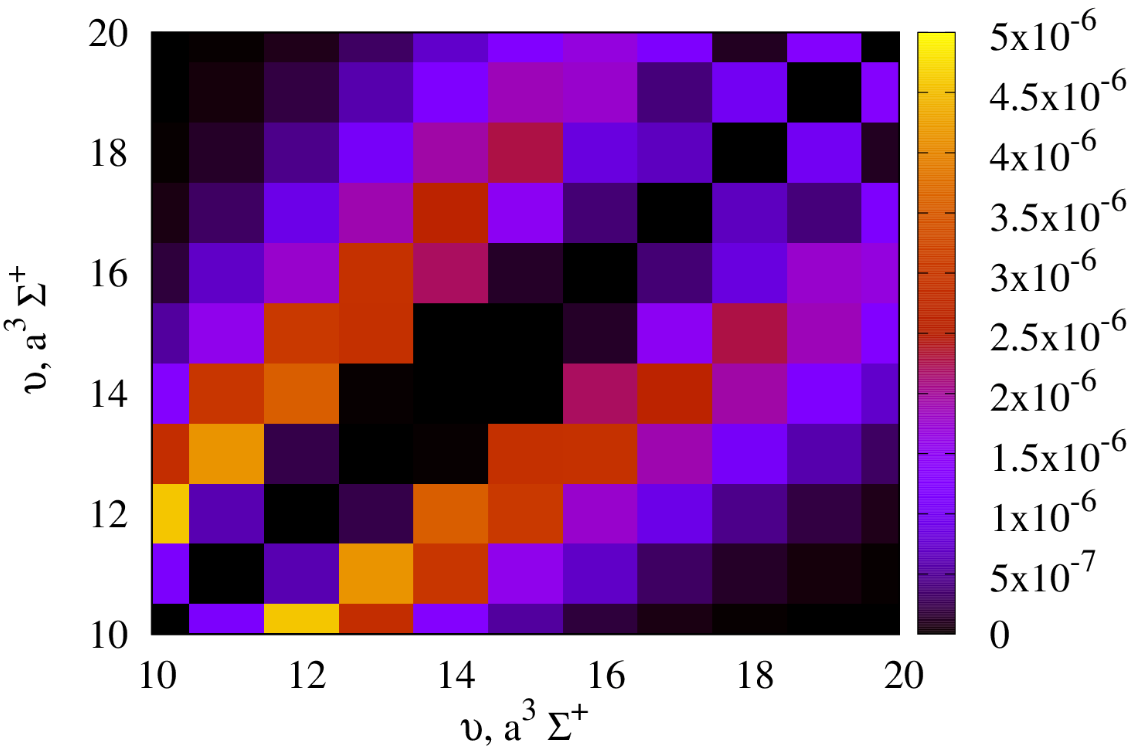}
\caption{}
\end{subfigure}
\caption{(Color online) (a) Squared dipole matrix element map (in atomic units) of the $a^{3}\Sigma^{+}$ potential energy curve of KRb. (b) Zoom of the ``hole" around $\upsilon=14$.}
\label{fig:SDME}
\end{figure}

The idea behind our OLD scheme can be viewed as the reverse of the ladder-climbing scheme proposed by Chelkowski \emph{et al.} \cite{Chelkowski1990}: The molecule is successively driven from the initial level $\upsilon=i$ down the ladder  $i-1,i-2,\cdots,f$, where $\upsilon=f$ is the target level, employing a single LCP (see Fig. \ref{fig:LD_schemes}). We employ a QOC method to adjust the parameters that define the shape of this pulse so as to maximize the sequential population transfer between the ladder rungs.
This is possible in heteronuclear diatomic molecules where the permanent electric dipole moment, $D(R)$, provides a significant coupling between adjacent levels.
To assess such coupling, we examine the squared dipole matrix elements (SDMEs) within the $a^{3}\Sigma^{+}$ electronic state,
\begin{equation}\label{eq:SDME}
D_{\upsilon,\upsilon'}\equiv|\bra{\upsilon}D\ket{\upsilon'}|^{2}.
\end{equation}
Figure \ref{fig:SDME}(a) displays the corresponding SDME map.
It can be observed that, indeed, the coupling of the level $\upsilon=10$ with the levels $\upsilon>16$ is very weak.
Moreover, it can be seen that the dipole coupling between adjacent levels is relatively strong, except that around $\upsilon=14$ there is a ``hole" in the coupling map, as can be more clearly appreciated in Fig. \ref{fig:SDME}(b), where such hole is zoomed in. Since this can represent a problem for an OLD scheme, we also consider a MLD scheme, where this hole is skipped (see Fig. \ref{fig:LD_schemes}).


\section{\label{sec:method}Methodology}

We used the KRb $a^{3}\Sigma^{+}$ PEC, $V(R)$, and $D(R)$ reported in Ref. \citenum{Borsalino2014}.
We calculated the vibrational eigenenergies and eigenfunctions of the (nonrotating) $^{39}$K$^{87}$Rb isotopologue by numerical integration of the time-independent nuclear Schr\"odinger equation,
\begin{equation}\label{eq:TISE}
    \left[-\frac{\hbar^{2}}{2\mu}\frac{d^{2}}{dR^{2}} + V(R)- E_{\upsilon}\right] \psi_{\upsilon}(R)=0,
\end{equation}
where $\mu$ is the reduced mass of the nuclei, using a Colbert-Miller discrete variable representation (DVR) \cite{Colbert1992}.
Then, we evaluated the SDMEs (\ref{eq:SDME}) by numerical quadrature.

Within the semiclassical dipole approximation, the time-dependent nuclear Schr\"odinger equation takes the form
\begin{equation}\label{eq:TDSE}
    \left[-\frac{\hbar^{2}}{2\mu}\frac{d^{2}}{dR^{2}}+V(R)+\varepsilon(t)D(R)-i\hbar\frac{\partial}{dt}\right]\Psi(R,t)=0.
\end{equation}
where $\varepsilon(t)$ is the electric-field amplitude.
We integrated this equation numerically, representing the wave function on a space-time grid and approximating the short-time evolution operator by means of the midpoint quadrature and the symmetric Strang splitting,
\begin{equation}\label{STEO}
\hat{U}(t_n,t_{n+1})
\approx e^{-i\hat{T}\delta t/2\hbar}e^{-iW(\bar{t}_n)\delta t/\hbar}e^{-i\hat{T}\delta t/2\hbar},
\end{equation}
where $\delta t\equiv t_{n+1}-t_n$ is the time step, $\bar{t}_n\equiv t_n+\delta t/2$ is the midpoint time, $\hat{T}$ is the kinetic-energy operator, and $W(t)\equiv V(R)+\varepsilon(t)D(R)$ is the effective time-dependent potential. This approximation is accurate to $\cal{O}$$(\delta t^3)$ \cite{Guerrero2015}.

Since the interaction with the field can induce absorption above the dissociation threshold, besides stimulated emission, part of the wave function can escape into the continuum.
When the latter reaches the end of the grid, an artificial reflection occurs that introduces a spurious back-emission into the bound levels.
To avoid this effect, we added a complex absorbing potential (CAP) \cite{cap} in the asymptotic region, with the form
\begin{equation}\label{CAP}
    V_A(R)=-i\eta(R-R_{0})^{2},
\end{equation}
where $R_{0}$ is the grid point at which this potential starts acting.

The electric field of the LCP has the Gaussian shape \cite{Guerrero2015} 
\begin{eqnarray}\label{eq:LCP}
    \varepsilon(t)&=&\varepsilon_{0}\exp\left[ -\frac{(t-\tau_{0})^{2}}{2\tau^{2}}\right],\nonumber\\
    &\times&\cos\left[\omega_{0}(t-\tau_{0})+\frac{1}{2}C(t-\tau_{0})^{2}\right],
\end{eqnarray}
where $\varepsilon_{0}$ and $\omega_{0}$ are the central amplitude and frequency, $\tau_{0}$ and $\tau$ are the time shift and width, and $C=d\omega/dt$ is the chirp parameter, with $\omega(t)=\omega_{0}+C(t-\tau_{0})$ being the instantaneous frequency. In this particular case, the energy difference between successive levels down the ladder increases, hence $C>0$.

\begin{figure}
\includegraphics[width=1.03\linewidth]{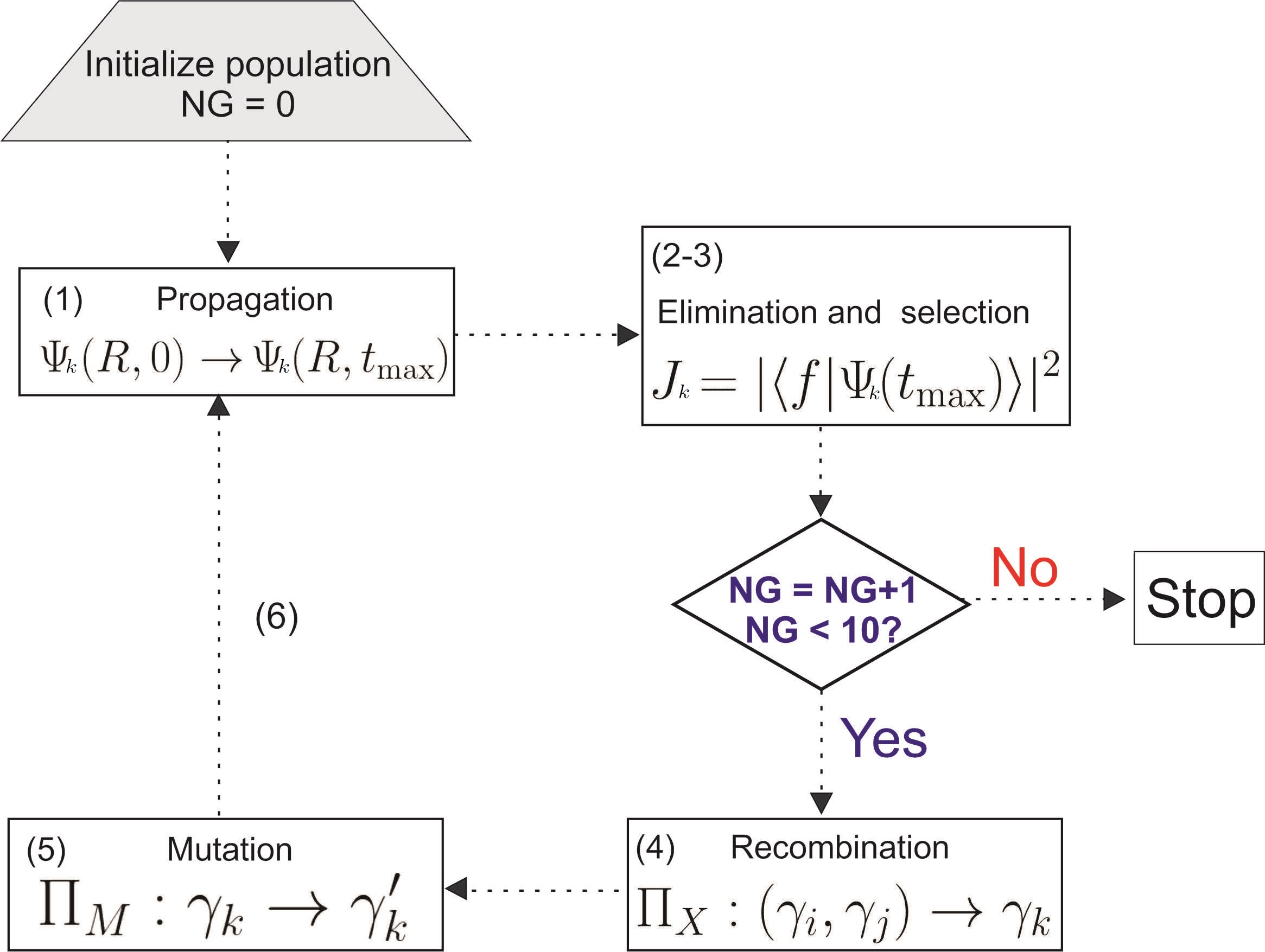}
\caption{Flow chart illustrating the steps involved in the genetic algorithm for pulse optimization. NG denotes the number of generations.}
\label{fig:chartGA}
\end{figure}

To find the optimal LCP parameters, we adapted the QOC+GA methodology of Arango and coworkers \cite{Guerrero2015,Guerrero2018}.
The $k$-th individual is a pulse whose chromosome consists of the 5-vector of genes $\gamma_{k}\equiv(\varepsilon_{0},\omega_0,\tau_{0},\tau,C)$.
According to the criteria explained below, we chose an initial population of 40 individuals ($k=1,\dots,40$), which constitute the generation zero.
Then, we evolved this population through the following optimization cycle (see Fig. \ref{fig:chartGA}):
(1) Propagation of the initial wave function $\psi_{\upsilon=i}(R)\equiv\Psi_k(R,0)\rightarrow\Psi_k(R,t_{max})$ with each one of the pulses separately, and calculation at $t_{max}$ of their scores with the fitness function and the cumulative fitness.
For the fitness function, we chose the survival probability of the target level,
\begin{equation}
J_k=|\bra{f}\ket{\Psi_k(t_{max})}|^{2}.
\end{equation}
Once the individuals are organized from lowest to highest fitness, the cumulative fitness for a given individual, $k$, is given by $\sum_{j=1}^{k} J_j/\sum_{j=1}^{40} J_j$, where the denominator is a normalization factor.
(2) Selection of the best 5 individuals using the roulette-wheel selection method based on the cumulative fitness. This method involves the generation of a random number between 0 and 1 and its comparison with the cumulative fitness value of each individual; the higher the cumulative fitness of an individual, the greater the chance of being selected.
(3) Elimination of the remaining 35 individuals from the population.
Here, if the number of generations is less than 10 we continue with step (4), otherwise we stop the optimization and choose the best individual.
(4) Replacement of the eliminated individuals with the children generated by crossing over the survivors' genes with probability $\Pi_{X}$.
(5) Sampling of the mutation probability of all genes of each individual of the new population, followed by mutation of those with probabilities less than $\Pi_{M}$. This operation is not applied to the individuals selected in step (3).
(6) Return to step (1).

To generate the initial population of LCPs, we chose the initial parameters randomly within appropriate ranges, which were determined heuristically appealing to physical considerations, as follows.
The spectral bandwidth of the LCP \cite{bandwidth}
\begin{equation}
\label{BW1}
\sigma=2\sqrt{2\ln 2}
\sqrt{\frac{1}{\tau^{2}}+\tau^2c^2},
\end{equation}
must include at least the range of frequencies required for the successive transitions, $\Delta \omega\equiv\omega_{f-1,f}-\omega_{i,i-1}$, where $\omega_{f-1,f}$ and $\omega_{i,i-1}$ are the frequencies of the last and first transitions, respectively, in the LD scheme.
We set $C\approx\Delta\omega/6\tau$, hence this bandwidth takes the form
\begin{equation}
\label{BW2}
\sigma\approx 2\sqrt{2\ln 2}\sqrt{\frac{1}{\tau^{2}}+\frac{\Delta\omega^{2}}{36}}.
\end{equation}
The condition $\sigma\sim\Delta\omega$ allows us to estimate a lower bound for $\tau$.
Then, we chose $\tau_{0}$ to be about three times the standard deviation of the Gaussian in Eq. (\ref{eq:LCP}).
We made sure that the values of $\tau$ and $\tau_0$ be much shorter than the radiative lifetime of the initial vibrational level with respect to spontaneous emission into the vibrational manifold of the $a^{3}\Sigma^{+}$ electronic state (the lifetimes of the lower-lying levels are longer), $\tau_{i}=\sum_{\upsilon<i}A_{i\upsilon}^{-1}$, where $A_{i\upsilon}=2\omega_{i\upsilon}^3 D_{i\upsilon}^2/3\epsilon_0 c^3\hbar$ is an Einstein coefficient, with $\omega_{i\upsilon}$ being a transition frequency.
Since $\omega(t=0)\approx\omega_{i,i-1}$, we get that $\omega_{0}\approx\omega_{i,i-1}+\tau_0\Delta\omega/6\tau$.

To determine the range of $\varepsilon_0$ we took into account that the Rabi period for any of the sequential transitions is $T\sim\left(\bar{\varepsilon}D_{\upsilon,\upsilon'}\right)^{-1}$, where $\bar{\varepsilon}$ is the mean amplitude of the pulse during the transition, and that the range of $T$ must be consistent with the range of $\tau$.
The resulting value of $\varepsilon_0$ must not be too high, to avoid ionization of the molecule.

To perform the numerical integration of Eq. (\ref{eq:TDSE}), coupled with the QOC+GA pulse optimization, we adapted the computer code \cite{code} employed in Ref. \citenum{Guerrero2018}. 

\section{\label{sec:results}Results and Discussion}

For the numerical integration of Eqs. (\ref{eq:TISE}) and (\ref{eq:TDSE}), we obtained converged results employing a grid of 140 bohr and 5600 grid points, and a CAP with $R_{0}=100$ bohr and $\eta=5\times10^{-6}$ hartree/bohr$^2$.
We found 30 bound vibrational levels in the $a^3\Sigma^+$ electronic state.
We obtained that the lifetime of the highest-lying vibrational level with respect to spontaneous emission into the vibrational manifold of the $a^{3}\Sigma^{+}$ electronic state is $\tau_{\upsilon=30}\approx 13$ s.

After a few trials, we determined that suitable values for the GA probabilities are $\Pi_{X}=0.25$ and $\Pi_{M}=0.9$.

For the initial states $i=20,24$ we obtained the condition $\tau>3.17\times 10^{5}$ atomic units $=7.7$ ps.

\subsection{\label{subsec:OLD}One-rung-at-a-time Ladder Descending}

Table \ref{tab:OLDparameters} presents the initial ranges, chosen in accordance with the criteria explained in Sec. \ref{sec:method}, and the optimal values, yielded by the GA methodology, of the LCP parameters. (Note that an optimal value may lie outside its initial range, which is an indication of the flexibility of the algorithm. The same observation applies to Table \ref{tab:MLDparameters} below.)
The optimal amplitudes turned out to be of the same order of magnitude as the ones reported in Ref. \citenum{Ndong2010}.
\begin{table}[!htbp]
\begin{ruledtabular}
\begin{center}
\caption{\label{tab:OLDparameters} Ranges of the LCP parameters for the GA optimization and optimal values obtained in the OLD scheme. All quantities are given in atomic units.}
\begin{tabular}{l c c c}
\multicolumn{4}{c}{$i=20$} \\
\hline
 & min  & max & optimal\\
\hline
$\tau$ & $1.0\times 10^6$ & $1.0\times 10^7$ & $9.798\times 10^6$\\
$\tau_0$ & $3.3\times 10^6$ & $3.5\times 10^7$ & $4.104\times 10^7$\\
$C$ & $4.0\times 10^{-13}$ & $5.0\times 10^{-12}$ & $6.259\times 10^{-13}$\\
$\omega_0$ & $3.1\times 10^{-5}$ & $3.6\times 10^{-5}$ & $3.531\times 10^{-5}$\\
$\varepsilon_0$ & $1.0\times 10^{-3}$ & $1.0\times 10^{-2}$ & $8.011\times 10^{-3}$\\
\hline
\multicolumn{4}{c}{$i=24$} \\
\hline
 & min  & max & optimal\\
\hline
$\tau$ & $1.0\times 10^6$ & $1.0\times 10^7$ & $1.146\times 10^7$\\
$\tau_0$ & $3.3\times 10^6$ & $3.5\times 10^7$ & $3.723\times 10^7$\\
$C$ & $6.0\times 10^{-13}$ & $7.0\times 10^{-12}$ & $7.300\times 10^{-13}$\\
$\omega_0$ & $3.3\times 10^{-5}$ & $3.6\times 10^{-5}$ & $3.723\times 10^{-5}$\\
$\varepsilon_0$ & $1.0\times 10^{-3}$ & $1.0\times 10^{-2}$ & $9.168\times 10^{-3}$\\
\end{tabular}
\end{center}
\end{ruledtabular}
\end{table}

\begin{figure}
\includegraphics[width=0.9\linewidth]{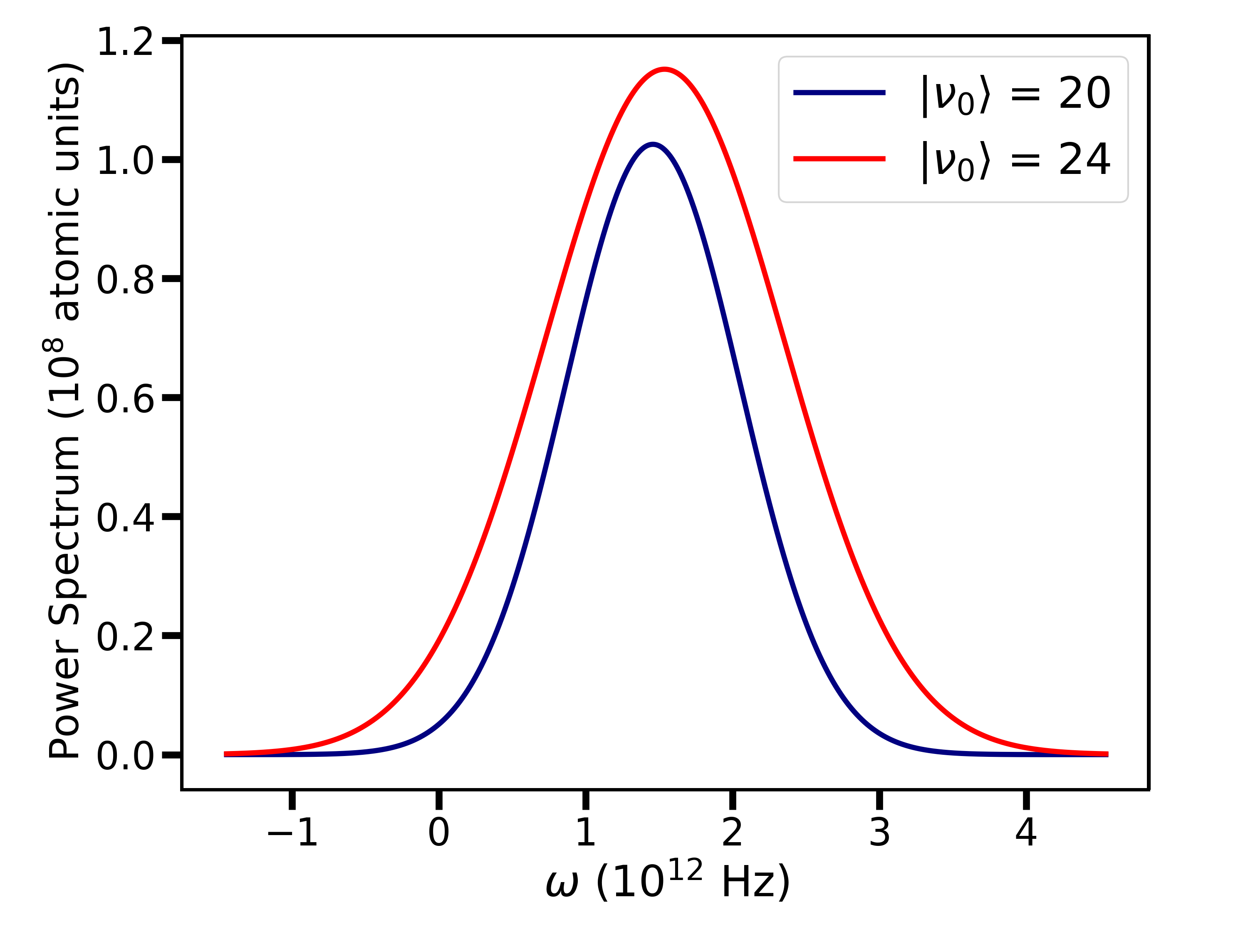}
\caption{(Color online) Optical spectra of the optimal pulses in the OLD scheme for the two initial levels.}
\label{fig:OLDspectrum}
\end{figure}

The optical spectrum of the LCP is given by
\begin{equation}\label{eq:spectrum}
I(\omega)=\sqrt{\frac{\tau^{4}}{1+c^{2}\tau^{4}}}\varepsilon_{0}^{2}\exp\left[-\frac{(\omega-\omega_{0})^{2}}{2\sigma^{2}}\right].
\end{equation}
Figure \ref{fig:OLDspectrum} displays the optical spectra of the LCPs for the two initial levels.
It is seen that the range of excitation frequencies is of the order of $10^{12}$ Hz, that is, in the infrared region, as expected for vibrational transitions.

\begin{figure}
\begin{subfigure}[h]{1.1\linewidth}
\includegraphics[width=0.80\linewidth]{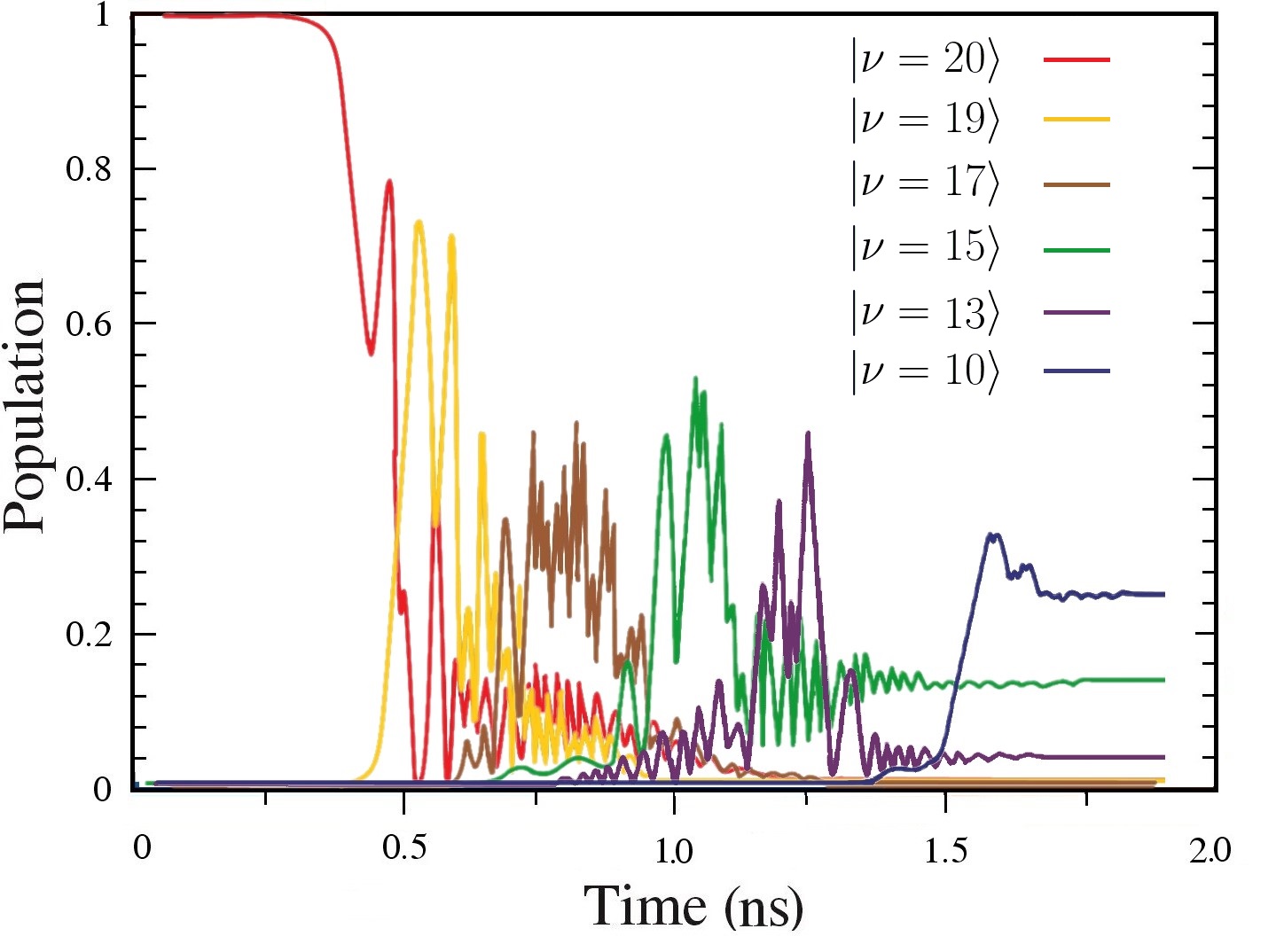}
\caption{}
\end{subfigure}
\begin{subfigure}[h]{1.02\linewidth}
\includegraphics[width=1.0\linewidth]{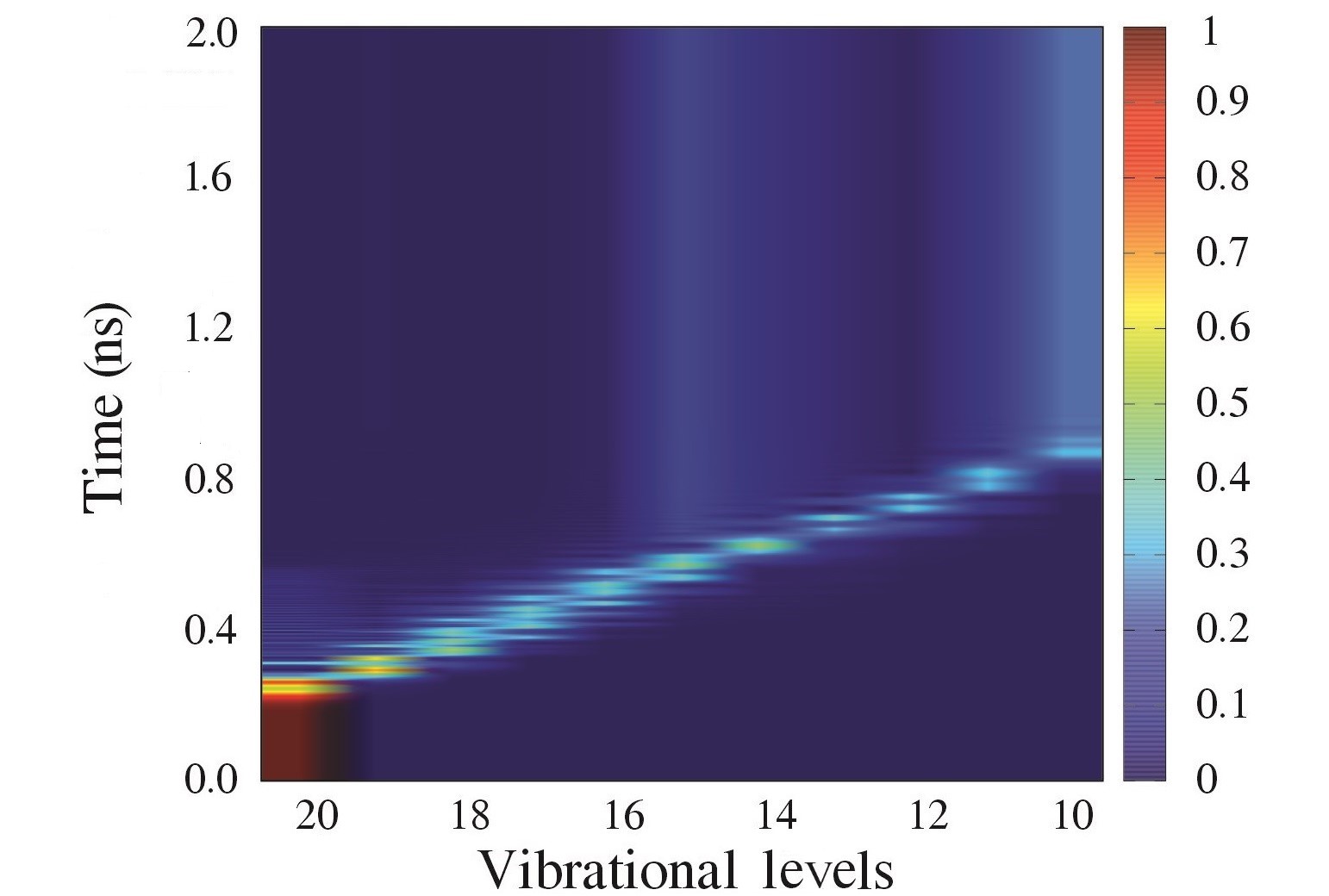}
\caption{}
\end{subfigure}%
\caption{(Color online) Time evolution of the populations in levels $10\leq\upsilon\leq 20$ in the OLD scheme when the initial level is $\upsilon=20$. (a) Selected levels. (b) All levels.}
\label{fig:OLD20}
\end{figure}

Figure \ref{fig:OLD20} shows the populations
\begin{equation}\label{eq:pop}
p_{\upsilon}(t)=|\bra{\upsilon}\ket{\Psi(t)}|^2,
\end{equation}
for the case where the initial level is $\upsilon=20$.
It is observed that, once the pulse begins to act on the system, the population is transferred down the ladder of levels in an approximately sequential manner.
Naturally, the transfer between any pair of levels cannot be complete, since the pulse amplitude and chirped frequency cannot fulfill exactly the conditions required for a full population transfer in a two-level system.
However, the initial level is totally emptied before the pulse is over.

At the end of the pulse, the population in the target level is $p_{\upsilon=10}=25\%$, while a large portion of the remaining population remains in level $\upsilon=15$. In Fig. \ref{fig:SDME} it can be appreciated that this level is at the edge of the hole in the SDME map, thus causing a bottleneck for the population transfer towards lower-lying levels.

The final populations in the bound levels add up to only 55\%. The population loss is attributed to the aforementioned dissociation that results from absorption, especially at the early stages of the molecule-field interaction. We will discuss this phenomenon in more detail in Sec. \ref{subsec:MLD}.

\begin{figure}
\begin{subfigure}[h]{1.1\linewidth}
\includegraphics[width=0.80\linewidth]{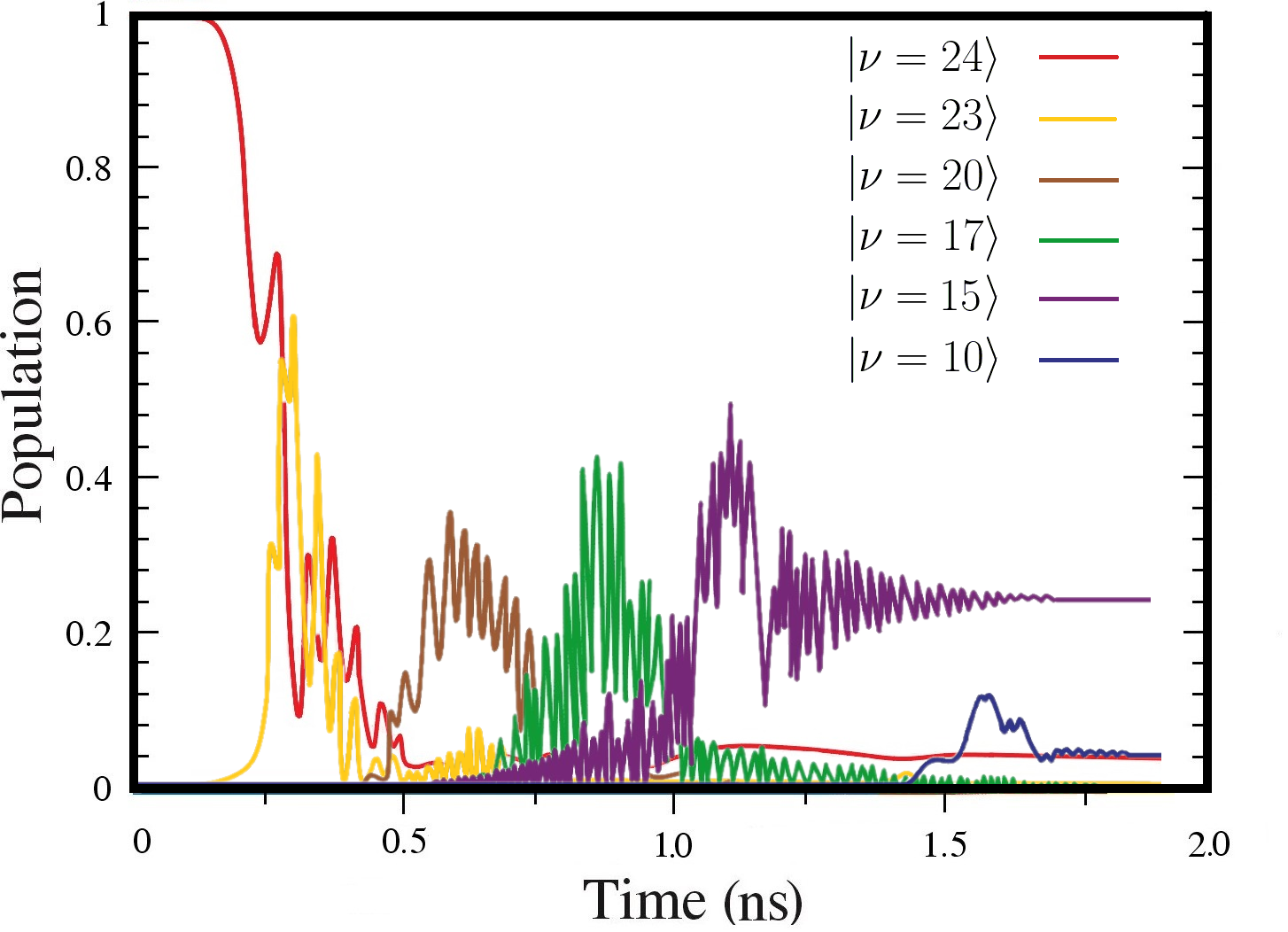}
\caption{}
\end{subfigure}
\begin{subfigure}[h]{1.02\linewidth}
\includegraphics[width=0.93\linewidth]{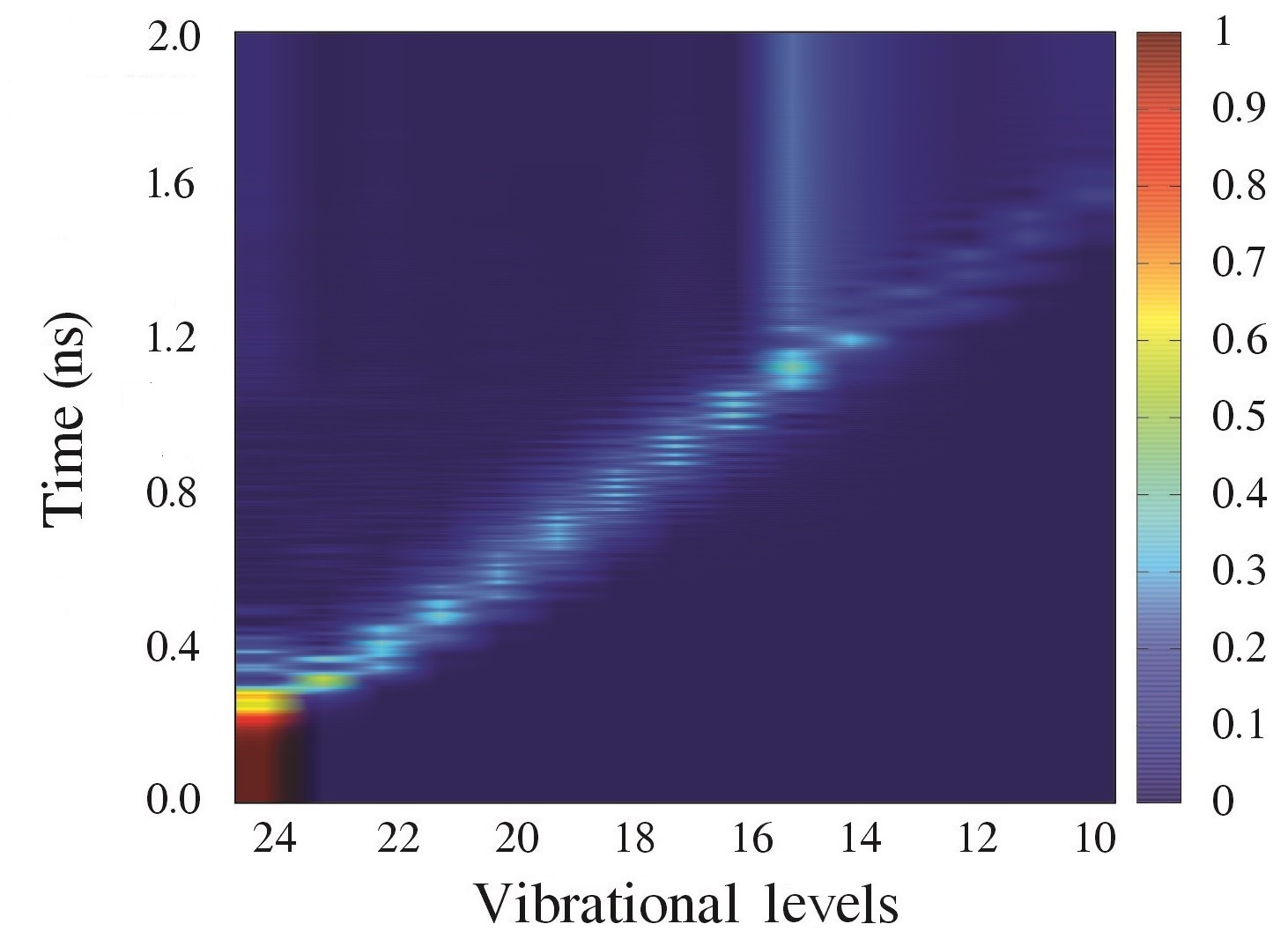}
\caption{}
\end{subfigure}%
\caption{(Color online) Time evolution of the populations in levels $10\leq\upsilon\leq 24$ in the OLD scheme when the initial level is $\upsilon=24$. (a) Selected levels. (b) All levels.}
\label{fig:OLD24}
\end{figure}

Figure \ref{fig:OLD24} displays the populations for the case where the initial level is $\upsilon=24$. The LD mechanism is very clear until level $\upsilon=15$ is reached, when the bottleneck is strongly manifested, causing the population in the target level at the end of the pulse to be only $p_{\upsilon=10}=5\%$, while $p_{\upsilon=15}=23\%$.
The total population in the bound levels is 45\%. The population lost to dissociation is now higher, as the initial level is closer to the dissociation threshold.

The explanation of the marked difference in the two cases is the following.
When the initial level is $\upsilon=20$, the system must climb down 10 levels to reach the target level, $\upsilon=10$. Hence, the center of the hole, $\upsilon=14$, is almost at the middle of the vibrational ladder, which is reached when the pulse amplitude is at its maximum, $\varepsilon(\tau_{0})$.
Consequently, the low coupling between adjacent levels within the hole is compensated by the high field amplitude, permitting a significant population transfer before the chirp takes the field out of resonance. 
On the other hand, when the initial level is $\upsilon=24$ the system must climb down 14 levels to reach the target level, and this matching cannot occur.
Such mismatch could be mitigated by tailoring asymmetric pulse shapes, but this would complicate the optimization and, even worse, the experimental implementation of the LD scheme.
Therefore, we next explore a strategy where the system is made to ``jump over the hole" still using a Gaussian pulse shape.

\subsection{\label{subsec:MLD}Multiple-rung-at-a-time Ladder Descending}

For the MLD variant we chose four (non-adjacent) levels, skipping the hole at $\upsilon=14$.
These must satisfy two conditions: subsequent levels exhibit a relatively strong coupling and the energy differences increase down the ladder, so that $C>0$.
(The levels could be chosen so that the energy differences decrease and $C<0$, but this would imply that the energy difference between the first two levels had to be relatively large, which in turn would imply that the dipole coupling for the first transition would be relatively weak, thereby probably rendering the process to be less efficient.)
The selected transitions and their energies for both initial states are shown in Table \ref{table:transitions}.
\begin{table}[!htbp]
\begin{ruledtabular}
\begin{center}
\caption{\label{table:transitions} Selected transitions in the MLD scheme for the two initial states. Energy differences are given in atomic units.}
\begin{tabular}{cccc}
\multicolumn{2}{c}{$\upsilon_{0}=20$} & \multicolumn{2}{c}{$\upsilon_{0}=24$}\\
\hline
\textrm{$\upsilon\longrightarrow \upsilon'$}&
\textrm{$\Delta$E}&
\textrm{$\upsilon\longrightarrow \upsilon'$}&
\textrm{$\Delta$E}\\
\hline
20 $\longrightarrow$ 16 & 1.068 $\times$ 10$^{-4}$ &24 $\longrightarrow$ 17 & 1.372 $\times$10$^{-4}$\\
16 $\longrightarrow$ 13 & 1.110 $\times$ 10$^{-4}$ & 17 $\longrightarrow$ 13 & 1.423 $\times$ 10$^{-4}$\\
13 $\longrightarrow$ 10 & 1.378 $\times$ 10$^{-4}$ & 13 $\longrightarrow$ 10  & 1.430 $\times $10$^{-4}$ \\
\end{tabular}
\end{center}
\end{ruledtabular}
\end{table}

Table \ref{tab:MLDparameters} displays the initial ranges and the optimal values of the LCP parameters.
The optimal amplitudes turned out to be somewhat lower than the ones of the OLD variant.
Since $\omega_0$ for the MLD scheme is an order of magnitude larger than the one for the OLD scheme, the time scale of the MLD process is one order of magnitude shorter than the one for the OLD process, as revealed by the values of $\tau$, $\tau_0$, and $C$.
\begin{table}[!htbp]
\begin{ruledtabular}
\begin{center}
\caption{\label{tab:MLDparameters} Ranges of the LCP parameters for the GA optimization and optimal values obtained in the MLD scheme. All quantities are given in atomic units.}
\begin{tabular}{l c c c}
\multicolumn{4}{c}{$i=20$} \\
\hline
 & min  & max & optimal\\
\hline
$\tau$ & $3.2\times 10^5$ & $3.2\times 10^6$ & $1.489\times 10^6$\\
$\tau_0$ & $1.0\times 10^6$ & $1.0\times 10^7$ & $4.900\times 10^6$\\
$C$ & $1.8\times 10^{-12}$ & $1.6\times 10^{-11}$ & $8.254\times 10^{-12}$\\
$\omega_0$ & $1.0\times 10^{-4}$ & $1.8\times 10^{-4}$ & $1.211\times 10^{-4}$\\
$\varepsilon_0$ & $1.0\times 10^{-3}$ & $1.0\times 10^{-2}$ & $5.154\times 10^{-3}$\\
\hline
\multicolumn{4}{c}{$i=24$} \\
\hline
 & min  & max & optimal\\
\hline
$\tau$ & $1.0\times 10^6$ & $1.0\times 10^7$ & $1.003\times 10^6$\\
$\tau_0$ & $3.3\times 10^6$ & $3.5\times 10^7$ & $4.835\times 10^6$\\
$C$ & $1.0\times 10^{-13}$ & $1.0\times 10^{-12}$ & $5.832\times 10^{-12}$\\
$\omega_0$ & $1.3\times 10^{-4}$ & $1.6\times 10^{-4}$ & $1.378\times 10^{-4}$\\
$\varepsilon_0$ & $1.0\times 10^{-3}$ & $1.0\times 10^{-2}$ & $5.720\times 10^{-3}$\\
\end{tabular}
\end{center}
\end{ruledtabular}
\end{table}

\begin{figure}
 \includegraphics[width=0.9\linewidth]{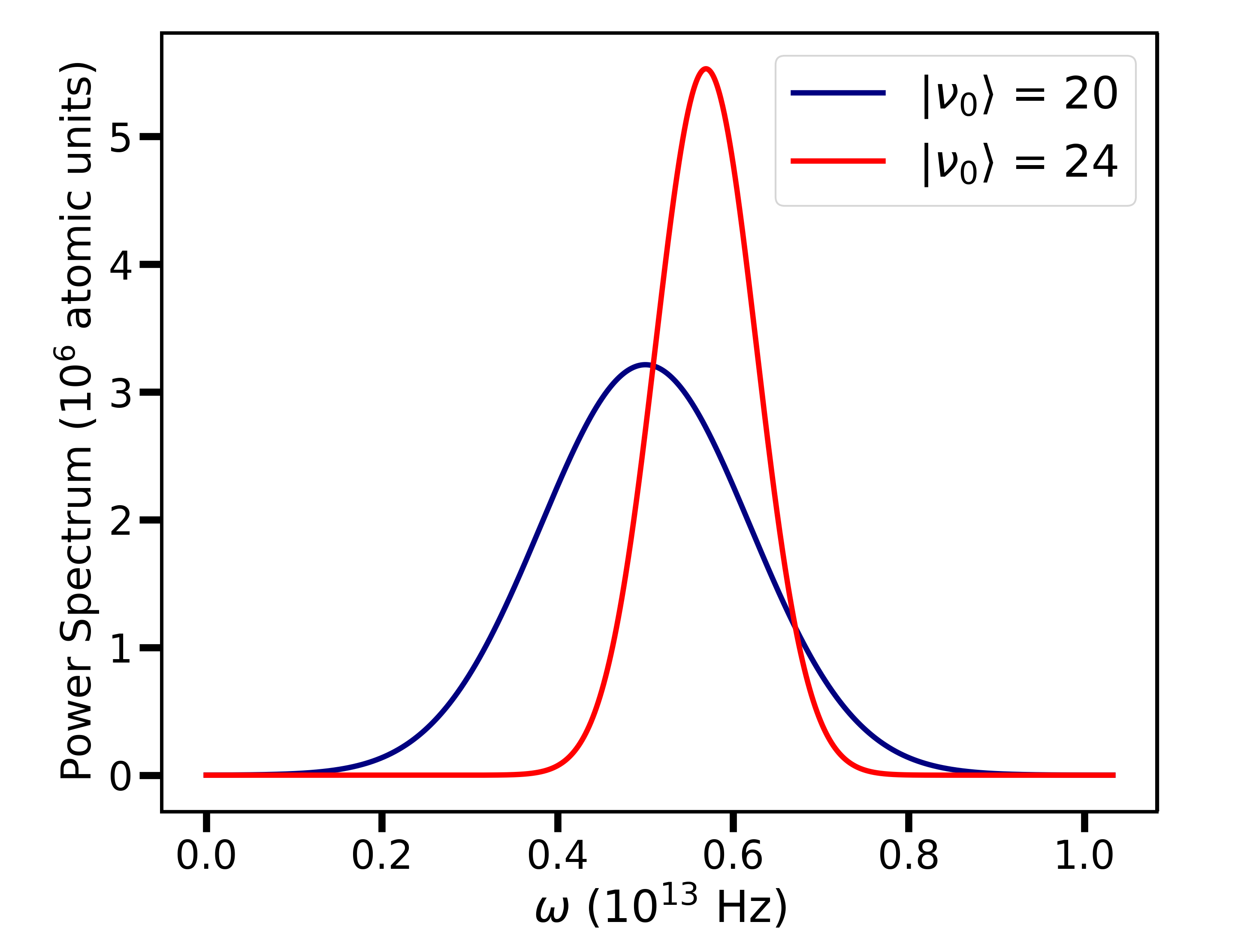}
\caption{(Color online) Optical spectra of the optimal pulses in the MLD scheme for the two initial levels.}
\label{fig:MLDspectrum}
\end{figure}

Figure \ref{fig:MLDspectrum} shows the optical spectra of the LCPs for the two initial levels.
It can be appreciated that the range of excited
frequencies is shifted towards higher values in comparison with the ones of the OLD scheme, but still is within the infrared region.

\begin{figure}
\begin{subfigure}[h]{1.1\linewidth}
\includegraphics[width=0.74\linewidth]{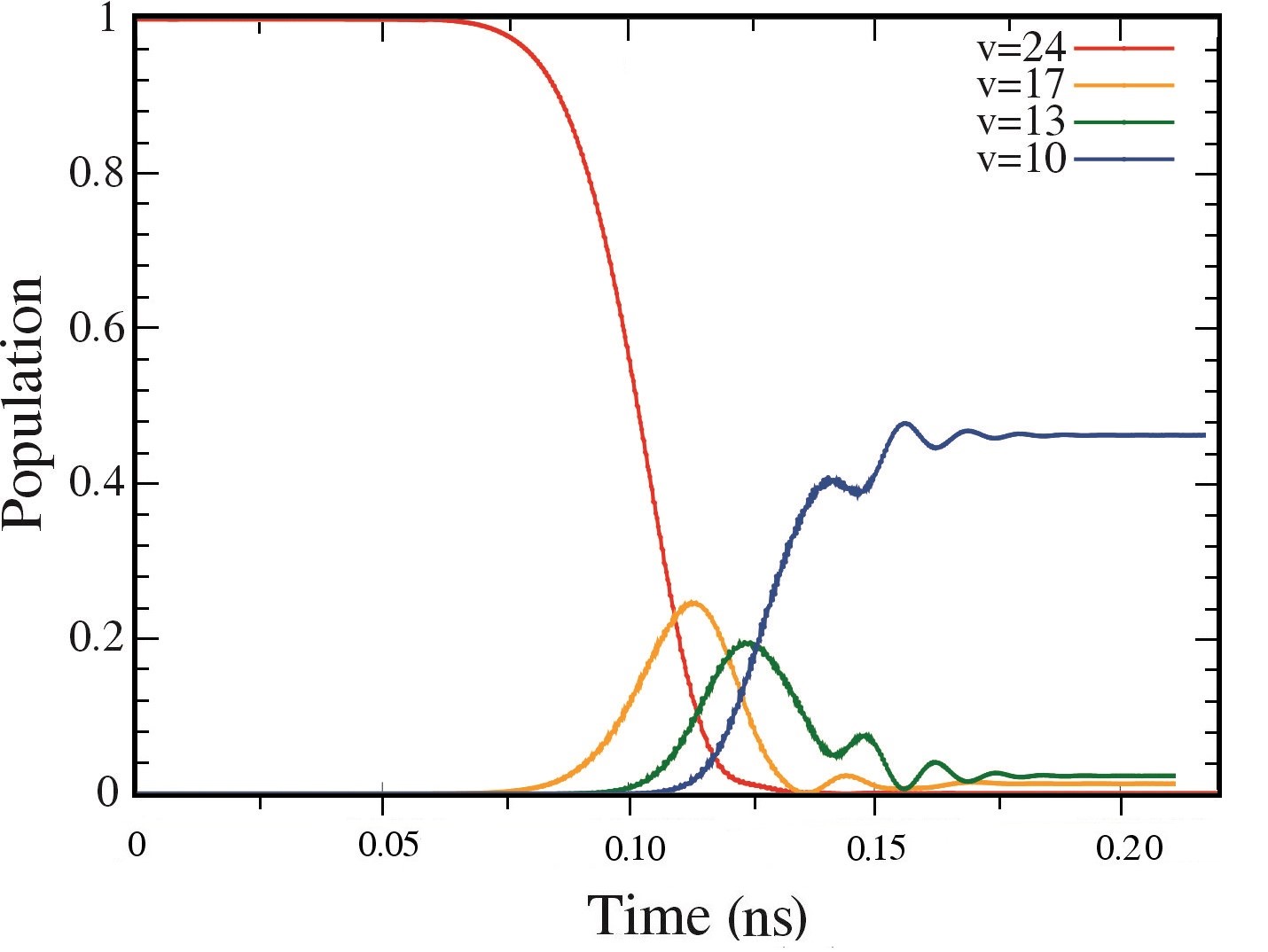}
\caption{}
\end{subfigure}
\begin{subfigure}[h]{1.02\linewidth}
\includegraphics[width=0.94\linewidth]{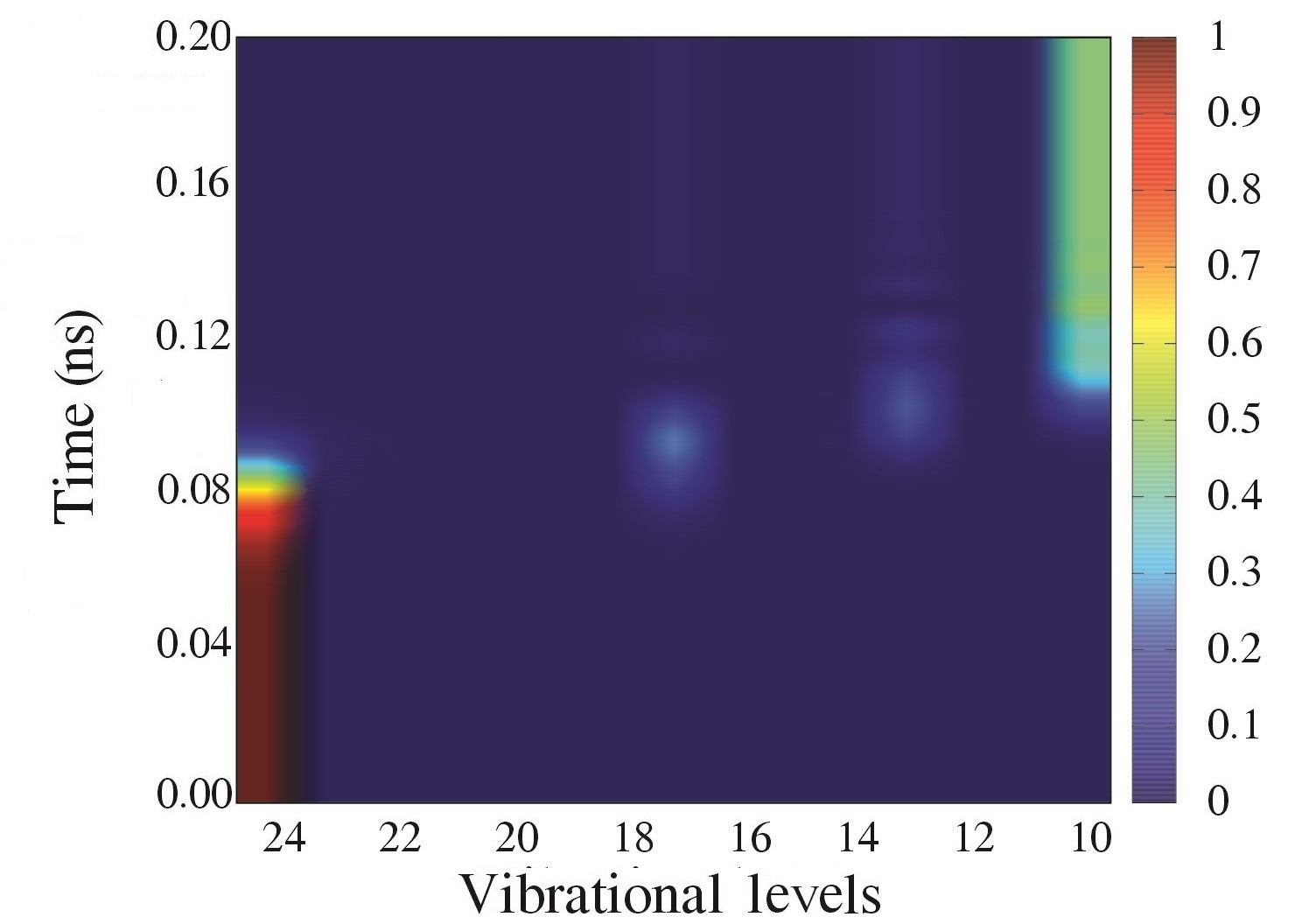}
\caption{}
\end{subfigure}
\caption{\label{fig:MLD24}(Color online) Time evolution of the level populations in the MLD scheme when the initial level is $\upsilon=24$. (a) Chosen levels. (b) All levels.}
\end{figure}

Figure \ref{fig:MLD24} displays the population dynamics for the case where the initial level is $\upsilon=24$. The LD nature of the process is evident. The final population in the target level is $p_{\upsilon=10}\approx 48\%$, which amounts to an increase of $43\%$ with respect to the OLD scheme.
At no time during the process is population transferred to bound levels not explicitly included in the scheme.
The process takes about 0.175 ns, versus about 1.6 ns in the OLD scheme, an order of magnitude shorter, as pointed out above.
The high-frequency oscillations have practically disappeared, which is a signature of the decrease in the antiresonant contributions caused by the increase of $\omega_0$ and decrease of $\varepsilon_{0}$, as can be inferred, for example, from the familiar expression provided by time-dependent perturbation theory for the transition amplitude, which contains the denominators $\omega_0+\omega_{\upsilon,\upsilon'}$.

\begin{figure}[h]
\centering
\includegraphics[width=0.44\textwidth]{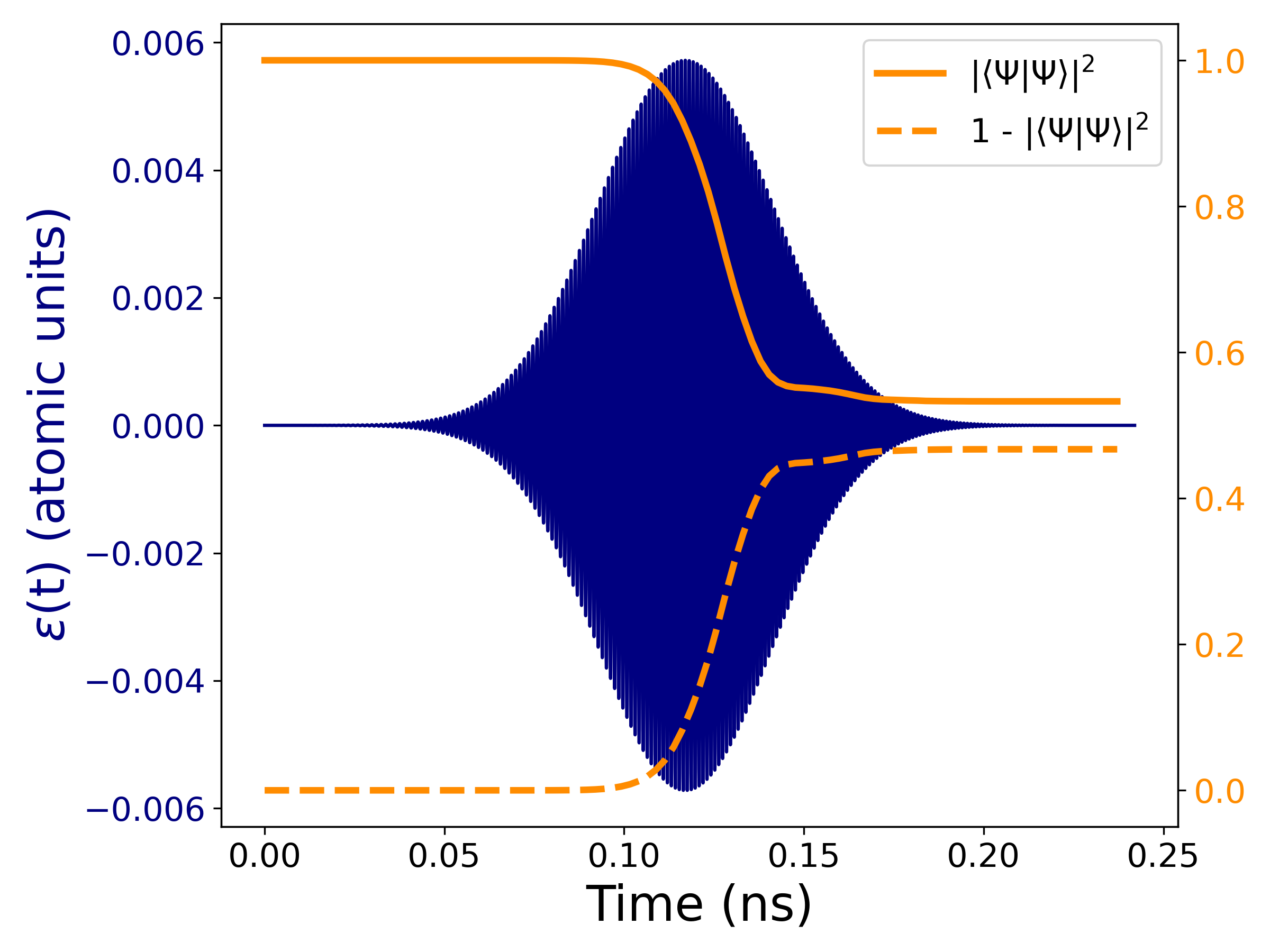}
\caption{(Color online) Time evolution of the optimal field amplitude and of the total bound and continuum populations in the MLD scheme when the initial level is $\upsilon=24$.}  
\label{fig:dissociation}
\end{figure}

The final populations in the bound levels add up to 52\%, which means that almost all the \emph{bound} population was transferred to the target, the remaining 4\% of the bound population residing in the rest of the levels. But Fig. \ref{fig:MLD24} shows that nearly all of this 4\% resides in levels lying below the initial one. The missing 48\% of the total population must have been lost to dissociation.
To verify this conclusion, Fig. \ref{fig:dissociation} shows the total probability ($\bra{\Psi(t)}\ket{\Psi(t)}^2$) and the dissociation probability ($1-\bra{\Psi(t)}\ket{\Psi(t)}^2$) superimposed on the time-dependent field amplitude.
It is observed that at about 0.1 ns the total probability begins to decrease, as the CAP begins damping the continuum part of the wave function. The delay of about 0.05 ns with respect to the beginning of the pulse is the time taken by such part of the wave function to propagate to the CAP region.
At about 0.18 ns all the continuum part of the wave function has been absorbed, and the remaining probability lies in the bound levels, which, indeed, amounts to approximately 48\%.
Naturally, the dissociation probability mirrors the total probability.

\begin{figure}
\begin{subfigure}[h]{1.1\linewidth}
\includegraphics[width=0.80\linewidth]{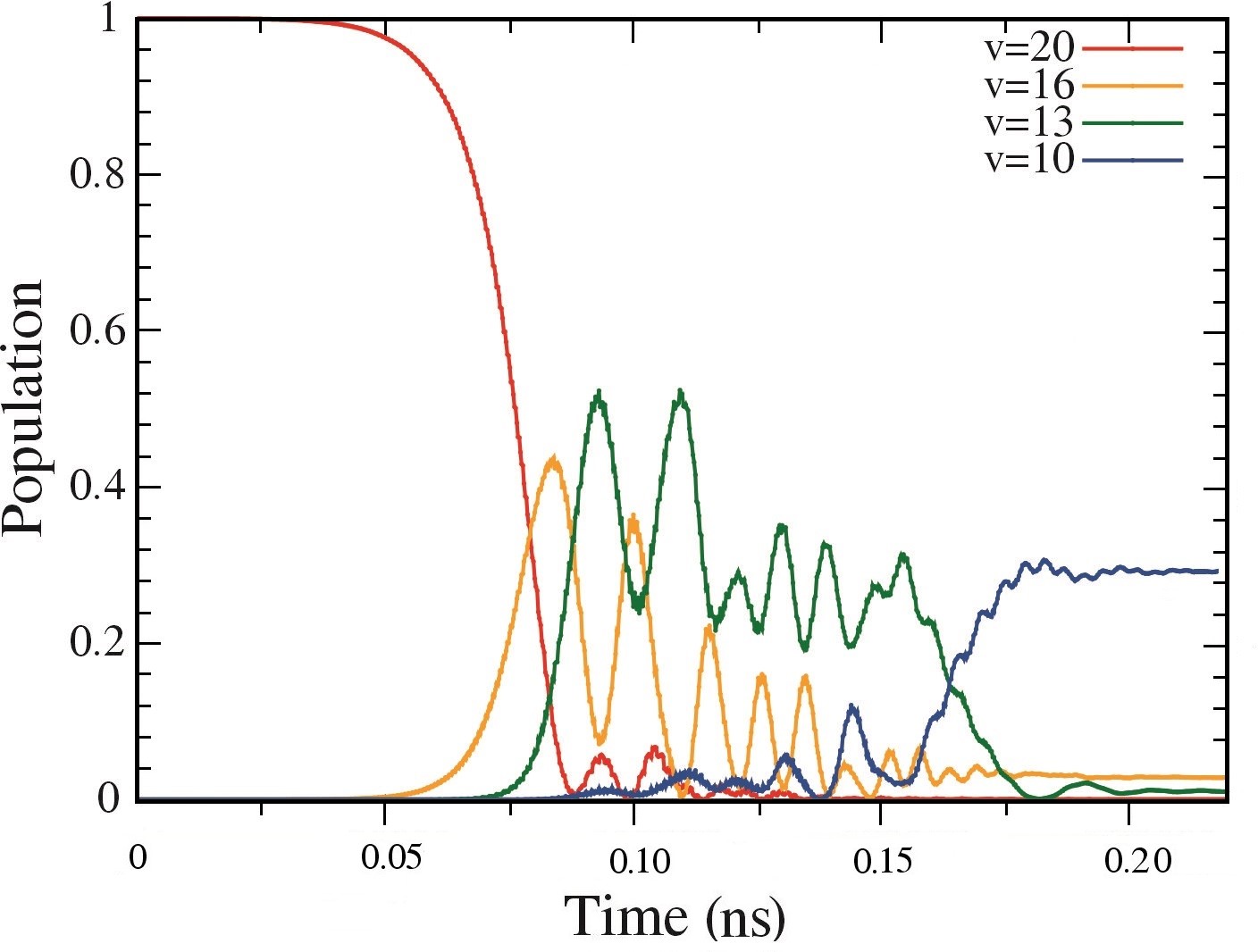}
\caption{}
\end{subfigure}
\begin{subfigure}[h]{1.02\linewidth}
\includegraphics[width=0.96\linewidth]{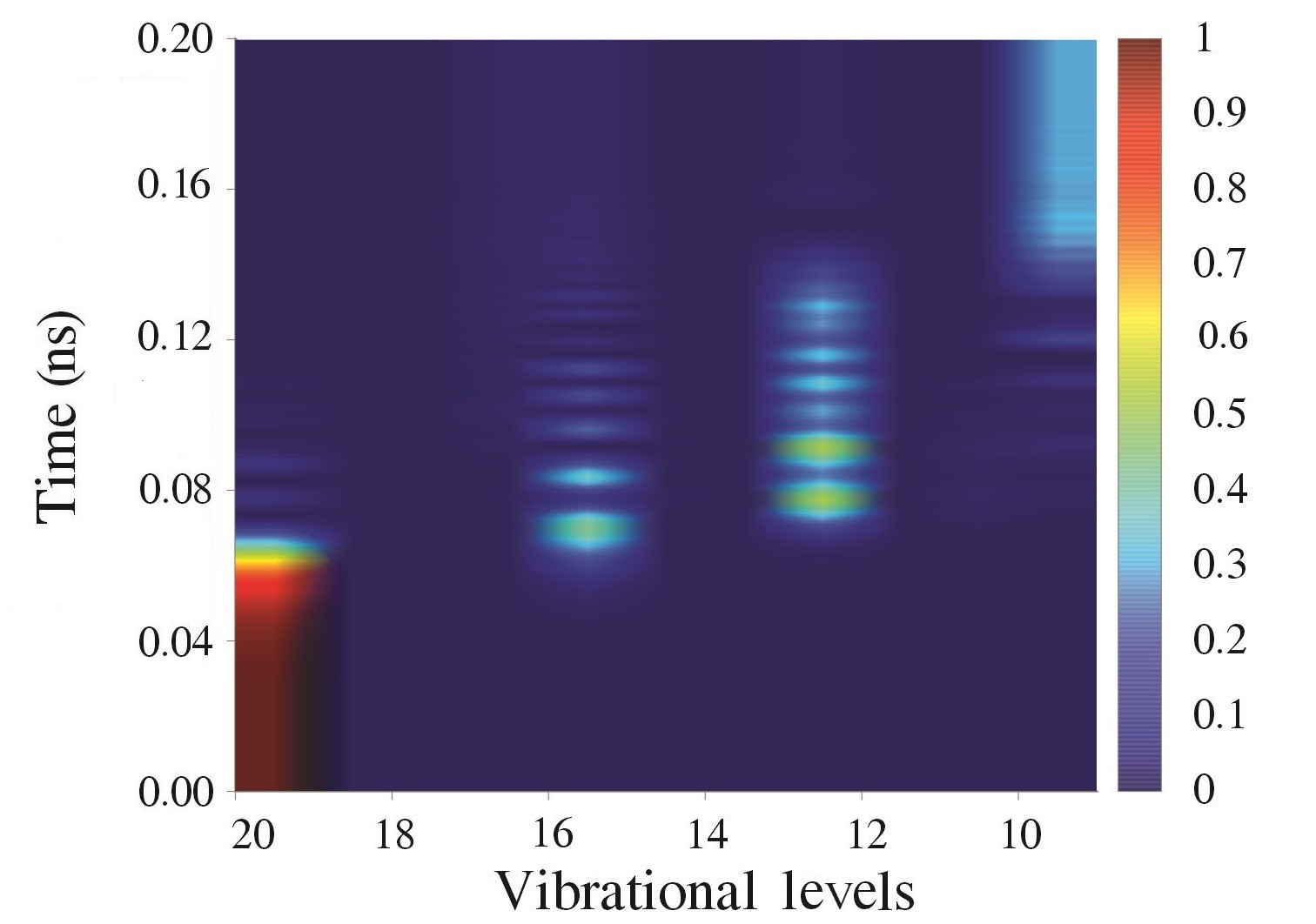}
\caption{}
\end{subfigure}
\caption{\label{fig:MLD20}(Color online) Time evolution of the level populations in the MLD scheme when the initial level is $\upsilon=20$. (a) Chosen levels. (b) All levels.}
\end{figure}

Figure \ref{fig:MLD20} shows the population dynamics for the case where the initial level is $\upsilon=20$.
Now, complementary transient Rabi oscillations in the $\upsilon=16$ and $\upsilon=13$ populations, lasting for about 0.075 ns and with a middle time of about $0.12$ ns $\approx\tau_0$, are clearly exhibited, indicating that the dynamics get temporarily stuck in this two-level system, although the LD nature of the process can still be appreciated.
This comes about because when the field frequency sweeps through the energy difference between those two levels the field amplitude is at its maximum, making the Rabi frequency sufficiently high for several oscillations to occur while the two levels are near resonance. Besides, the  populations of the other levels are very small around this time, causing little interference.
As this two-level system decays, the population of the target level rises to a final value of $p_{\upsilon=10}=30\%$, which amounts to an increase of only $5\%$ with respect to the OLD scheme.
This happens because by the time the target level is reached, the field amplitude is already too low.
The total population in the bound levels is 40\%.
This case illustrates that the selection of the levels constituting the ladder is crucial for the efficiency of the MLD scheme.

As in Ref. \citenum{Guerrero2018}, we have assumed that the molecule is isolated and the dynamics are fully coherent. Hence, we have neglected radiative decay and temperature-dependent effects, namely, vibrational relaxation and population re-thermalization induced by black-body radiation.
We found that the radiative lifetimes of the vibrational levels with respect to spontaneous emission within the same electronic state ($a^{3}\Sigma^{+}$) are longer than 13 s, which turns out to be much longer than the duration of the entire LD process, which takes at most a few nanoseconds. Moreover, spontaneous emission into the electronic ground state ($X^{1}\Sigma^{+}$) is spin-forbidden. Therefore, it is a good approximation to neglect radiative decay altogether.
At cold temperatures and sufficiently low gas densities, vibrational relaxation time scales can be much longer than nanoseconds.\cite{Forrey1999}
Likewise, at these temperatures population re-thermalization induced by black-body radiation takes of the order of seconds.\cite{Leibfried2012}
Hence, it is valid to neglect these two effects for our infrared LD processes.

The experimental realization of the proposed schemes necessitates subnanosecond laser pulses in the far-infrared range.
The generation of ultrashort laser pulses in the 10-40 THz spectral range has been achieved by the excitation of high harmonics in the organic nonlinear optical crystal DAST \cite{Kuroda2010, Manikandan2015}. Furthermore, pulses in the spectral range of those shown in Fig. \ref{fig:MLDspectrum} could be generated by employing novel quantum cascade molecular lasers, as demonstrated in Ref. \citenum{Mammez2021}. We expect that, in the near future, even lower frequencies in the THz spectral range, needed to implement our scheme in general, will be achieved by designing new nonlinear crystals \cite{ Sun2017} or by the exploitation of available lasing transitions in molecules.\cite{Fan2021}

\section{\label{sec:conclusions} Conclusions and Outlook}

We have proposed and implemented computationally an infrared ladder-descending scheme for the stabilization of a highly excited polar diatomic molecule into a given target vibrational level of the same electronic state. The scheme employs a single linear chirped laser pulse with an analytical shape that is optimized by means of a quantum optimal control method based on a genetic algorithm.
The implementation requires some heuristics based on the vibrational level structure and the dipole coupling map of the molecule.
This vibrational stabilization scheme can be used as a ``post-pulse" for varios types of association methodologies \cite{Juarros2006,Kotochigova2007,Marquetand2007,Kallush2008,Molano2019,Ulmanis2012,deLima2017,Kohler2006,Castano2020} or as a ``pre-pulse" for further optimization of pump-dump \cite{Sage2005,Guerrero2018} or STIRAP \cite{Devolder2021,Aikawa2010,Borsalino2014} stabilization methodologies.

To prove our concept, as a prototype we considered a model KRb molecule formed by magnetoassociation in its lowest-lying triplet electronic state, $a^{3}\Sigma^{+}$ \cite{Guerrero2018}.
This molecule exhibits a ``hole" in the dipole coupling map that can generate a bottleneck for a one-rung-at-a-time descent down the vibrational ladder for some initial levels.
We demonstrated that such bottleneck can be sidestepped by means of a multiple-rung-at-a-time variant of the scheme, taking advantage of the relatively strong overtones present.
Other molecules, bialkali or otherwise, may exhibit more complicated features in their dipole coupling maps, for example several holes. It seems that the multiple-rung-at-a-time variant of our scheme can deal with these cases by a judicious choice of the rungs, i.e., of the levels involved in the process.

We employed a Gaussian shape for the laser pulse, which is relatively easy to achieve experimentally. Our methodology can accommodate a more flexible shape, but at the obvious expense of increasing the optimization cost and the experimental difficulty.
For the genetic selection operation, we used the roulette wheel selection method. It would be worthwhile to try other selection methods that might improve the efficiency of the optimization.

The model employed does not take into account the rotational structure of the diatomic molecule. However, our previous study of the one-step photoassociation dynamics, which took into account the full rovibrational structure, revealed that for each vibrational level the rotational population distribution can become considerably wide, beyond what could be expected from the $\Delta J=\pm 1$ one-photon selection rule, due to the multiphoton character of the transitions when the field is sufficiently strong \cite{Molano2019}. Nevertheless, we expect that our multiple-rung-at-a-time scheme can achieve simultaneous one-step photoassociation and rovibrational stabilization, by hand-picking rovibrational levels in such a way that the rung separations increase (or decrease), so that a positive (or negative) frequency chirp can be employed. Work in these directions is currently underway in our laboratory. 



\begin{acknowledgments}

We are grateful to Diego F. Uribe and Javier Madro\~nero for useful discussions,
and to the Solid State Theory Group of Universidad del Valle for kindly providing time on their computing facilities.
This work was supported in part by Colciencias through Project No. 1106-658-42793.

\end{acknowledgments}

\newpage
\bibliography{Londono_ladderdescending}

\end{document}